# Using High-Speed WANs and Network Data Caches to Enable Remote and Distributed Visualization

Wes Bethel[1], Brian Tierney[2], Jason Lee[2], Dan Gunter[2], Stephen Lau[1]

*Lawrence Berkeley National Laboratory*

*University of California, Berkeley*

*Berkeley, CA 94720*

## 1.0 Abstract

Visapult is a prototype application and framework for remote visualization of large scientific datasets. We approach the technical challenges of tera-scale visualization with a unique architecture that employs high speed WANs and network data caches for data staging and transmission. This architecture allows for the use of available cache and compute resources at arbitrary locations on the network. High data throughput rates and network utilization are achieved by parallelizing I/O at each stage in the application, and by pipelining the visualization process. On the desktop, the graphics interactivity is effectively decoupled from the latency inherent in network applications. We present a detailed performance analysis of the application, and improvements resulting from field-test analysis conducted as part of the DOE Combustion Corridor project.

## 2.0 Introduction

As computing power increases, scientific simulations and instruments grow in size and complexity, resulting in a corresponding increase in output. During recent years, the increases in speed of infrastructure components that must absorb this output, including storage systems, networks and visualization engines, has not paced the increases in processor speeds. In response, solutions have tended toward parallel aggregations of slower, serial components, such as file systems striped across disk units.

**FIGURE 1. Visualization and Rendering Pipeline**

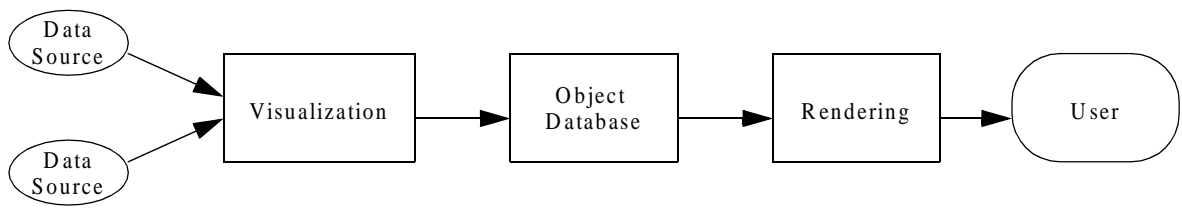

In particular, visualization and rendering pose interesting challenges as data sizes increase. In the visualization and rendering pipeline (Figure 1), abstract scientific data is first transformed into renderable data, such as geometry and image-based data, through the process of visualization. The resultant, renderable data is then transformed into a viewable image by a "draw" or rendering process. The challenges posed by large-model visualization stem from the sheer size of the data; it often won't fit within the confines of primary or secondary storage on a typical desktop workstation. Movement of large amounts of data to the workstation over typical network links is impractically slow, but even if practical, the graphics systems of even high-end workstations quickly become overwhelmed.

---

1. [ewbethel, slau]@lbl.gov,Visualization Group.
2. [bltierney, jrlee, dkgunter]@lbl.gov, Distributed and Data Intensive Computing Group.

Traditionally, visualization of large models has been approached using one of two strategies. In the first strategy, which we'll call "render remote," images are created on a large machine, preferably the same machine that has direct access to the data source (local filesystem), then transmitted to the user who views them on a workstation. In Figure 1, the link between *Rendering* and *User* would be over a network connection. In this configuration, a high-capacity resource has the potential to be applied to larger-sized problems than could be addressed with desktop resources, but graphics interactivity suffers due to the combination of latency and high bandwidth requirements[3]. In the second strategy, which we'll call "render local," smaller portions of the data, subsets or decimated versions of the raw data, are sent to the workstation where visualization and rendering take place. The network connection in this case is between *Data Source* and *Visualization.* Increasing graphics capacity mitigates concerns about interactivity, but the constraints encountered when moving remote data to the local workstation are exacerbated by limited network bandwidth and local storage capacity.

In recent years, two key developments have motivated us to explore a slightly different approach. One development is a network data cache that is tuned for wide-area network access, called the Distributed Parallel Storage System [1], or DPSS. The DPSS is a scalable, high-performance, distributed-parallel data storage system developed at Lawrence Berkeley National Laboratory (LBL). The DPSS is a data block server, built using low-cost commodity hardware components and custom software to provide parallelism at the disk, server, and network level. This technology has been quite successful in providing an economical, high-performance, widely distributed, and highly scalable architecture for caching large amounts of data that may potentially be used by many different users. Current performance results are 980 Mbps across a LAN and 570 Mbps across a WAN.

The other key development is a proliferation of high-speed, testbed networks. There are currently a number of Next Generation Internet networks whose goal is to provide network speeds of 100 or more times the current speed of the Internet. These include NSF's Abilene [2], DARPA's Supernet [3], and the ESnet testbeds [4]. Sites connected to these networks typically have WAN connection at speeds of OC12 (622 Mbps) or OC48 (2.4 Gbps); speeds that are greater than most local area networks (LANs). Access to these networks enables new options for remote, distributed visualization.

The combined capabilities of emerging high speed networks and scalable network storage makes it possible to consider remote, distributed scientific visualization from a new perspective, one which combines the best of both traditional methods.

## 3.0  Visapult: A Remote, Distributed Visualization Application Prototype

The Visapult application and framework consists of two distributed components (Figure 2): a viewer and a back end. In the following sections, we discuss the architecture of these components. The rendering portion of the viewer is built upon a scene graph model that proves useful for both asynchronous updates, as well as acting as a framework for the display of divergent types of data. The back end is a parallelized software volume rendering engine that uses a domain-decomposed partitioning, including the capability to perform parallel read operations over the network to a storage cache as well as parallel I/O to the viewer. Together, the viewer and back end implement a novel form of volume visualization that is fast but effective. More importantly, this novel form of volume visualization has been completely parallelized through the visualization and rendering pipeline, from the data source to the display. We describe our use of the DPSS as a network storage cache, as well as our methodology for obtaining performance data from the application.

---

3. 1K by 1K, RGBA images at 30fps requires a sustained transfer rate of 960Mbps.

**FIGURE 2. Visapult Architecture**

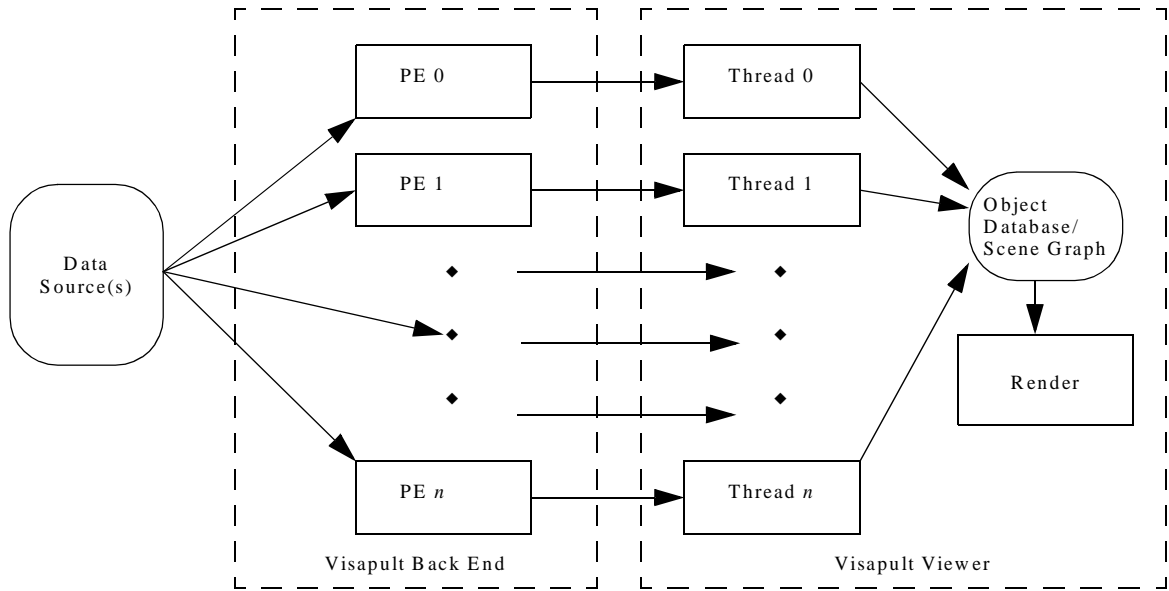

## 3.1 Visualization and Rendering Pipeline Architecture

The fundamental goal of Visapult, from a visualization perspective, is to provide the means to visualize and render large scientific data sets with interactive frame rates on the desktop or in an immersive virtual reality (VR) environment. In our design, we wanted the best of both worlds: performing as much visualization and rendering as possible on a parallel machine with either tera-scale data storage capacity, or a high-speed network link to such a storage resource, while leveraging the increasing graphics capacity of desktop and deskside workstations. A primary Visapult design goal, graphics interactivity, is a crucial, but subtle, part of the visualization process; studies have shown that motion parallax and a stereo display format increase cognitive understanding of three dimensional depth relationships by 200%, as compared to viewing the same data in a still image [7].

One troublesome dilemma is the speed difference between the infrastructure components and the problem size: disk transfer and network bandwidth rates are typically on the order of tens of megabytes per second, but data sizes are on the order of hundreds of gigabytes. How does one achieve interactivity on the desktop without moving all the data to the desktop?

Considering the visualization and rendering pipeline from Figure 1, we observe that in order to deploy a visualization tool on the desktop which is capable of rendering large data sets at interactive rates, the "object database" used by the renderer must be small enough to fit on the display platform. To that end, we have implemented a relatively new technique for volume rendering with a unique architecture that produces a relatively small object database, or scene graph[4]. As will be discussed later in the paper, we use a unique combination of task partitioning and parallelism to perform interactive volume visualization of large scientific data sets. Since visualization and rendering are pipelined and occur asynchronously, the viewer, which is "downstream" from the parallel software volume renderer, can interact with the render-

LBNL-45365    3

able objects at interactive rates. Updates of the scene graph through the visualization pipeline asynchronously from rendering, and occur at whatever rate the underlying infrastructure can provide.

A scene graph interface provides not only the means for parallel and asynchronous updates, but also an "umbrella" framework for rendering divergent data types. The scene graph system used in our implementation [8] supports storage and rendering of surface-based primitives (triangles, triangle strips, quads, polygons, etc.), vector-based primitives (lines, line strips), image-based data (volumes, textures, sprites and bitmaps), and text. The flexibility of this underlying infrastructure layer allows us to perform simultaneous rendering of volume and geometric data. Figure 3 is an image containing both volume rendering of density data, along with vector geometry (line segments) representing the adaptive grid created and used by the combustion simulation.

**FIGURE 3. Visapult Rendering of Combustion Data and Adaptive, Hierarchical Grids**

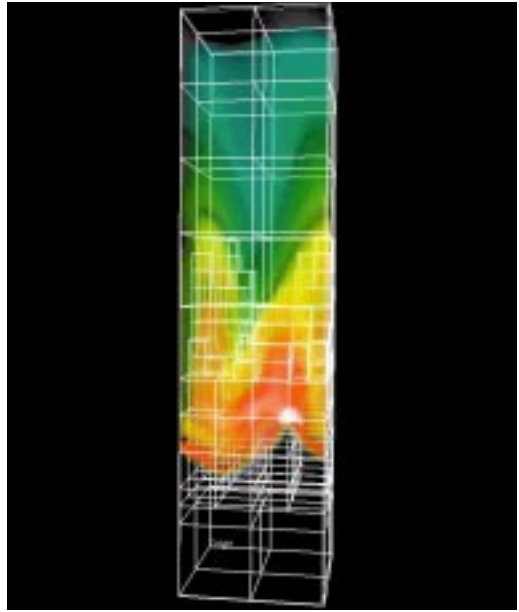

## 3.2 Parallel Volume Rendering Algorithm Taxonomy

Since volume rendering [9] is a computationally expensive and time consuming operation even with modest amounts of data, it is a likely candidate for parallelization. Algorithms for parallel volume rendering can be classified into two broad categories, *image order* and *object order,* based upon how the volume rendering task is decomposed across the pool of processors [10]. In an object order algorithm, the volume data is distributed across the processors using one of a number of different domain decomposition strategies (Figure 4). Each processor then renders its subset of the volume, producing an image. After all processors have finished rendering, the images from each processor must be gathered, then recombined into a final image. Recombination consists of image compositing using alpha blending [11], and must occur in a prescribed order (back-to-front or front-to-back). Note that each processor in an object order algorithm produces an intermediate image that may overlap in screen space with the images produced by other processors.

---

4. The term *scene graph* refers to a set of specialized data structures and associated services that provide management of displayable data and rendering services.

**FIGURE 4. Slab, Shaft and Block Decomposition**

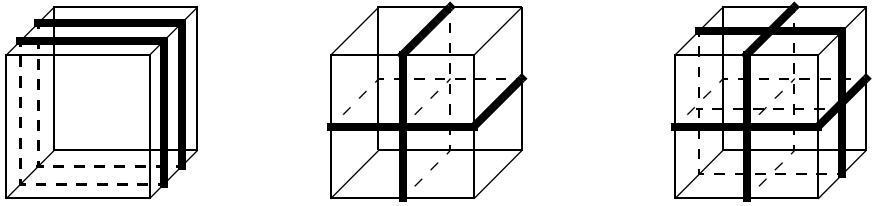

Image order algorithms, on the other hand, assign some region of screen space to each processor. The resulting images produced by each processor do not overlap, so recombination is not subject to an ordered image composition step. Depending upon the view, image order algorithms require some amount of data duplication across the processors, so do not scale as well with data size as the object order algorithms. The performance of image order parallel volume rendering algorithms is more sensitive to view orientation than the object order counterparts. In some views, there may be some processors with little or no work. In addition, as the model moves, the source volume data required at a given processor will change, requiring data redistribution as a function of model and view orientation.

## 3.3 Image Based Rendering Assisted Volume Rendering

Image based rendering (IBR) methods [12, 13] have been the subject of much attention in recent years. IBR methods are used primarily for generating different views of an environment from a set of pre-acquired imagery. The properties of IBR which make it attractive include interactive viewing with low computational cost irrespective of scene complexity, and the ability to use images from either digitized photographs or rendered models. Common among IBR methods is a process of warping and blending images from known views to represent what would be seen from an arbitrary view.

The concepts and principles of IBR model were recently applied to volume rendering [14]. Like the more conventional IBR counterparts, IBR assisted volume rendering (IBRAVR), seeks to achieve interactive rendering by avoiding the time-consuming process of completely rerendering the volume data for each frame. Instead, renderings of a model at arbitrary orientations are "computed" from "nearby" prerendered images. The prerendered images for the IBRAVR algorithm are obtained by volume rendering subsets of the entire volume. Using a slab decomposition, each source image would be obtained by volume rendering the slab of data. The total number of source images is equal to the number of data slabs created by data partitioning.

The per-frame, incremental rendering, or IBR component of IBRAVR, is implemented by using the pre-computed imagery as two dimensional textures which are texture-mapped onto geometry derived from the geometry of the slab decomposition, then rendered in depth order. In the basic algorithm, a single quadrilateral representing the center of the slab is used as the base geometry, and the computed imagery is texture mapped using alpha blending upon that geometry. With multiple slabs, there are multiple, overlapping, base geometries that are textured by the graphics hardware with the semi-transparent textures. As the model is rotated, the multiple textures correspondingly rotate in three dimensions, producing the impression of interactive volume rendering. As nearly all graphics hardware supports two-dimensional texturing, the IBRAVR viewer can be deployed on a wide variety of graphics platforms. An extension to this algorithm, described in [14], is replace the single quadrilateral with a quadrilateral mesh using offsets from the base plane for each point in the quad mesh. This enhancement will add a depth component to each of the IBR images, thereby enhancing the visualization process. We have included this extension in the Visapult implementation, but the details are omitted in this paper.

**FIGURE 5. IBR Assisted Volume Rendering**

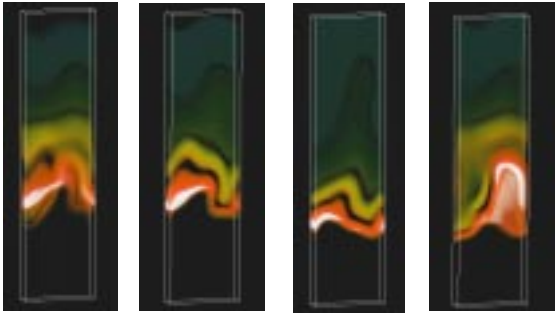

The source volume is subdivided into some number of slabs, each of which is volume rendered. The resulting images, along with geometric information derived from the original volume, are used as the source data for an IBR rendering engine.

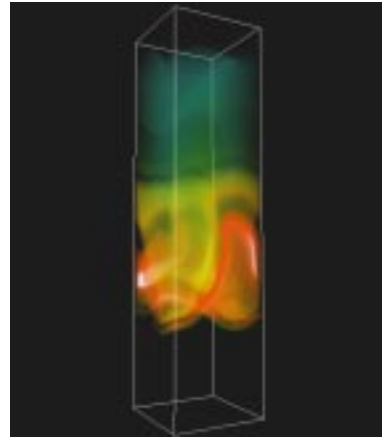

The final IBR model can be interactively transformed without the need to perform costly volume rendering on each frame.

As described in [14], the IBRAVR model exhibits visual artifacts as the model is rotated away from an axis-aligned view (Figure 6). These artifacts result from volume subdivision along an axis-aligned view, but rendered using a view or orientation that is not "closely" axis aligned. As the model rotates away from an axis-aligned view, the artifacts become more pronounced. [14] reports that objects viewed within a cone of about sixteen degrees will appear to be relatively free of visual artifacts.

**FIGURE 6. IBRAVR Artifacts**

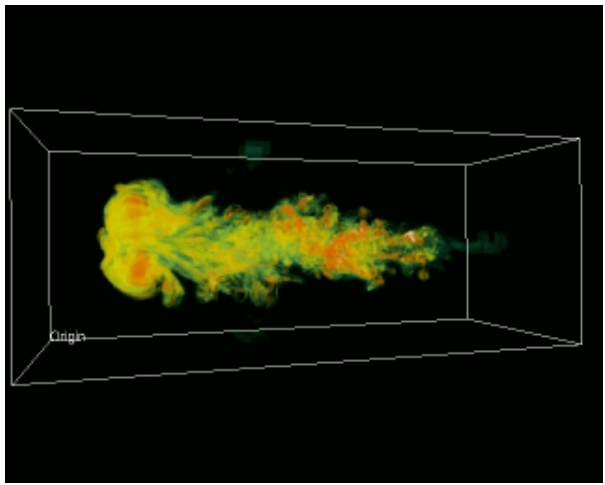
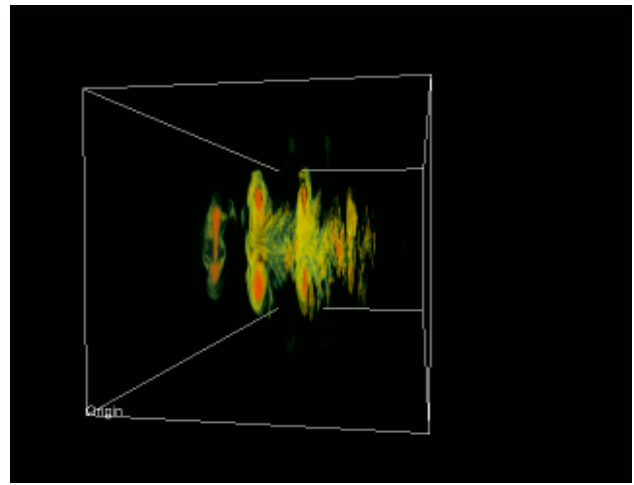

Using a nearly axis-aligned view, the IBRAVR method produces a high-fidelity image (left). When the model is rotated off-axis, visual artifacts can be seen (right). For the right image, we disabled axis-switching within Visapult, otherwise we would be viewing slices along the X-axis of the data.

Our implementation does not provide any remedies to this fundamental artifact of IBRAVR, but extends the base algorithm in a different manner that is useful for the purposes of visualization. On a per-frame basis, the Visapult viewer computes the best view axis, and transmits this information to the back end. The back end uses this information in order to select from either X-, Y-, or Z-axis aligned data slabs for use in volume rendering.

## 3.4 Visapult: Parallel and Remote IBRAVR

Visapult is a parallel and distributed implementation of an IBR assisted volume rendering engine. Our implementation can be thought of as a blend of an object-order parallel volume rendering engine with an IBRAVR viewer that uses a parallel, network-based data gathering model as an image assembly framework. The fundamental IBRAVR algorithm decomposes nicely into a distributed, pipelined and parallel architecture: a parallel object-order, parallel I/O capable volume rendering engine that produces images, and a parallel viewer that uses IBR techniques to assemble the individual images into a final display.

The Visapult back end reads raw scientific data from one of a number of different data sources, and each back end process performs volume rendering on some subset of the data, regardless of the viewpoint. The resulting images are transmitted to the Visapult viewer for final assembly into a model (scene graph), then rendered to the user. Owing to the IBRAVR design, the raw scientific data is distributed, or partitioned, amongst the back end processors using a slab-based decomposition (Figure 4). During the partitioning process, data is read into each processor in parallel. Each processor then performs software volume rendering upon its subset of the data. The resulting image from each processor is transmitted over the network to a peer receiver in the Visapult viewer, where it is inserted into the scene graph as a 2D texture.

On the viewer side, graphics interactivity results from a combination of the IBRAVR viewer model with a decoupling of scene graph updates from rendering. The amount of viewer-side data to be rendered is much smaller than the size of the raw volume data[5], so even software-only graphics systems are not overwhelmed. To implement the decoupling of rendering from scene graph updates, the viewer itself is a multi-threaded application, with one thread dedicated to interactive rendering, and other threads dedicated to receiving data from the Visapult back end visualization processes over multiple simultaneous network connections (implemented with a custom TCP-based protocol over striped sockets). Except for a small amount of scene graph access control with semaphores, I/O and rendering occur in an asynchronous fashion, so all pipes are full, making effective use of network and computational resources. Additional architectural details of the Visapult back end and viewer are presented in Appendix A.

## 3.5 Visapult's Use of the LBL DPSS as a Data Cache

In its role as data collector, the Visapult back end fetches raw scientific data for the purpose of visualization. One source of data is the DPSS, which is used as a storage cache for data sets that are too large to fit on the workstation. These data sets, generated on supercomputers or clusters of workstations, are typically on the order of 30 to 100 GB, and are often stored on archival systems such as HPSS [15], a high performance tertiary storage system. Clearly, it is impractical to transfer data sets of this magnitude to a local disk for processing. Also, archival systems such as the HPSS are not typically tuned for wide-area network access, and only provide full file, not block level, access to data. The DPSS addresses both of these issues; it is optimized for wide-area access to large files, and provides block level access, eliminating the need to transfer the entire file across the network. Therefore, we can migrate the files from HPSS to a nearby DPSS cache.

The DPSS provides several important and unique capabilities for data intensive distributed computing environments. It provides application-specific interfaces to an extremely large space of logical blocks. It offers the ability to build large, high-performance storage systems from inexpensive commodity components. It also offers the ability to increase performance by increasing the number of parallel disk servers.

---

5. Where the size of the raw volume data is $O(n^3)$, the amount of data to be rendered in the viewer is $O(n^2)$.

Typical DPSS implementations consist of several low-cost workstations as DPSS block servers, each with several disk controllers, and several disks on each controller. A four-server DPSS with a capacity of one Terabyte (costing about $15K in mid-2000) can thus deliver throughput of over 150 megabytes per second by providing parallel access to 15-20 disks. The overall architecture of the DPSS is illustrated in Figure 7.

**FIGURE 7. DPSS Architecture**

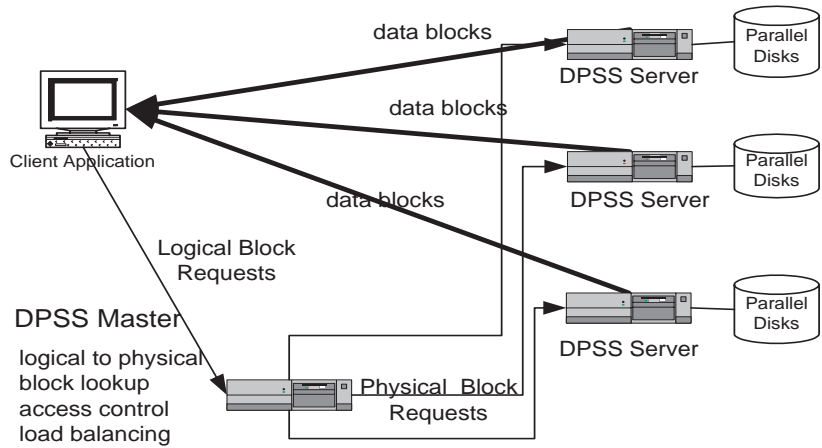

The application interface to the DPSS cache supports a variety of I/O semantics, including Unix-like I/O semantics, through an easy-to-use client API library (e.g., dpssOpen(), dpssRead(), dpssWrite(), dpssL-Seek(), dpssClose()). The DPSS client library is multi-threaded, where the number of client threads is equal to the number of DPSS servers. Therefore the speed of the client scales with the speed of the server, assuming the client host is powerful enough. This parallelism is leveraged by the parallel volume rendering performed by the Visapult back end.

## 3.6 Profiling and Performance Analysis - NetLogger

Profiling and analysis of an application's behavior and performance is an important part of the development process, but can prove challenging when the application consists of cooperative, distributed components. In our project, we made use of the NetLogger profiling toolkit for obtaining performance data from the application [16]. NetLogger includes tools for generating precision event logs that can be used to provide detailed end-to-end application and system level monitoring, and for visualizing log data to view the state of the distributed system. NetLogger has proven to be invaluable for diagnosing problems in networks and in distributed systems code. This approach is novel in that it combines network, host, and application-level monitoring, providing a complete view of the entire system.

The NetLogger system has a procedural interface: subroutine calls to generate NetLogger events are placed inside the source code of the application. Prior to running the application, a NetLogger daemon is launched on a host accessible to all components of the distributed application. During the course of application execution, the NetLogger subroutine calls communicate with the daemon host, where events are accumulated into an event log. This event log is then used as input for NetLogger visualization and analysis tools.

NLV, the NetLogger visualization tool, generates two dimensional plots from the raw data accumulated during a run. NetLogger and NLV were used extensively in Visapult field testing, and numerous examples of NLV output appear later in upcoming sections.

## 4.0 Visapult Field Testing and Evolution

In this section, we present several field testing experiments along with performance enhancements suggested by subsequent analysis. An early Visapult implementation was first presented at SC99 as part of a Research Exhibit. Since then, Visapult has become the reference application for a research program sponsored by the U.S. Department of Energy called *The Combustion Corridor*, and has been field-tested using several configurations of high speed testbed WANs using several different facilities. Research projects such as The Combustion Corridor seek to harness distributed resources for the purpose of scientific computing, such as high speed testbed networks, network storage systems, computational resources and large scale scientific data.

### 4.1 SC99 Research Exhibit

A preliminary version of Visapult was demonstrated at the SC99 conference in Portland, Oregon, reflecting a collaborative effort involving several research institutions: LBL, Sandia National Laboratory (SNL) and Argonne National Laboratory (ANL). Data from a cosmology hydrodynamic simulation[6] and a reactive chemistry combustion simulation[7] were transmitted over a WAN and visualized on the show floor at SC99. The demonstration required the use of NTON (National Transparent Optical Network) and SciNet99, the SC99 show floor network, to connect all of the resources (Figure 8).

**FIGURE 8. Visapult SC99 Configuration**

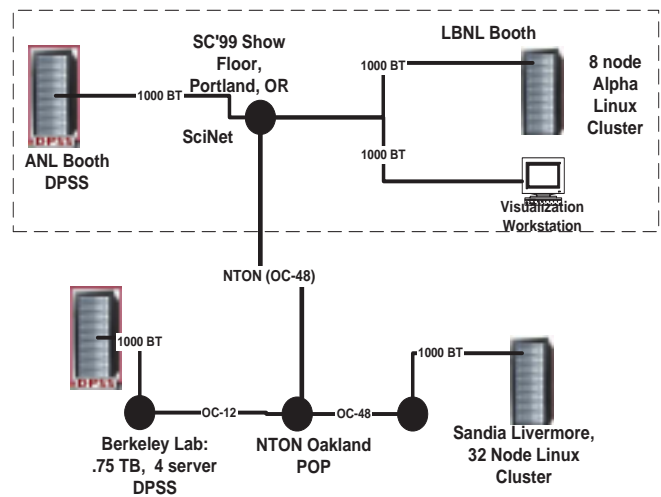

During the course of SC99, we used several different configurations of data sources, computational engines and networks as illustrated in Figure 8. Cosmology data was stored on DPSS systems at LBL and in the Argonne National Laboratory booth. Combustion data was stored on a parallel file system on the Cray T3E at the National Energy Research Scientific Computing Center (NERSC), located in Berkeley at LBL. Cosmology data was processed by a Visapult back end on the SNL CPlant [17] located in Livermore, California, or on the Babel Cluster [18] located in the LBL booth at SC99. The combustion data was pro-cessed by a Visapult back end running on the Cray T3E at NERSC in Berkeley. We also used multiple display devices for final rendering at SC99, including an ImmersaDesk located in the LBL booth, and a tiled surface display, located in the SNL booth. The ImmersaDesk allowed us to render the results in stereo. The tiled display system allowed us to demonstrate Visapult using a large-screen, theater-sized output format suitable for larger audiences.

---

6. Cosmology data courtesy of Julian Borrill, Scientific Computing Group, National Energy Research Scientific Computing Center (NERSC), Lawrence Berkeley National Laboratory.

7. Combustion data courtesy of Vince Beckner and John Bell, Center for Computational Sciences and Engineering, National Energy Research Scientific Computing Center (NERSC), Lawrence Berkeley National Laboratory.

We performed some preliminary analysis of the behavior of the system at SC99 using different network topologies and facilities. Our preliminary results showed that the majority of communication was between the DPSS (the network data cache) and the Visapult back end, with the link between the Visapult back end and viewer requiring much less bandwidth. This behavior is expected from the architecture of the system. Since the Visapult back end performs parallel volume visualization to reduce the data down to a small subset of images, it is expected that the amount of data resulting from the visualization and transmitted between the back end and viewer will be significantly less than the amount of data moved to the back end from the data source.

**FIGURE 9. Visualization of Hydrodynamic Cosmology Simulation Results at SC99**

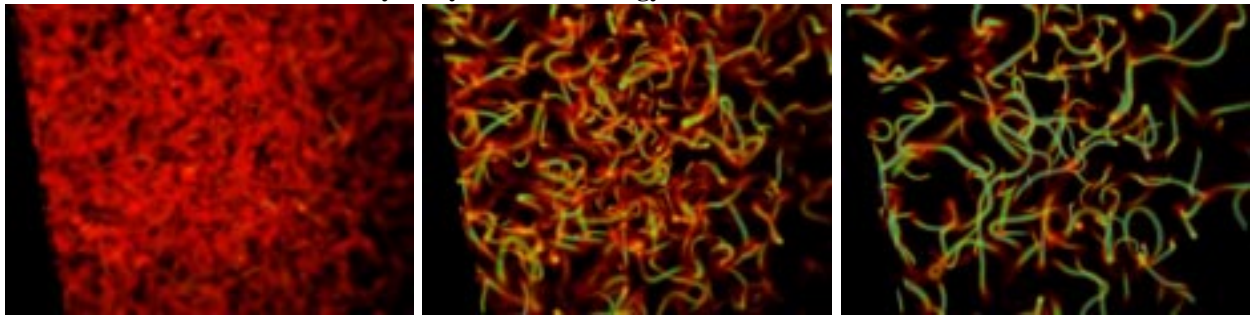

We were capable of sustaining a data transfer rate of 250Mbps between the DPSS located at LBL and CPlant, and a rate of 150Mbps between the DPSS at LBL and the LBL cluster at SC99. The difference in transfer rates was based upon the different network topologies. The link between the SC99 show floor and LBL required resource sharing over SciNet.

## 4.2 Combustion Corridor *First Light* Campaign

More recently, we have undertaken field testing using many of the same resources as for the SC99 project, but with an eye towards careful instrumentation and profiling analysis, and with larger data sets. This work is part of a project called *The Combustion Corridor*, sponsored by the U. S. Department of Energy, which is a collaborative research effort that includes LBL, ANL and SNL-CA. The term *Combustion Corridor* refers to the *process* of remote and collaborative visualization of large, scientific data sets for the Combustion Research community. The term "corridor" has been coined to refer to the metaphorical path from data source to human consumer, where the path spans geographical and system boundaries. A theme common across "corridor" projects is that many endeavors that were once possible only over LANs are now possible over WANs using a wider array of distributed resources. To a large extent, the needs and requirements of the Combustion Corridor are sufficiently general to be applicable to a wide variety of problem domains, including medicine, physics, and the geosciences.

Within the Combustion Corridor effort, we have performed several end-to-end runs using differing network topologies and platform configurations, which we refer to as "campaigns." The first such campaign took place on 12 April 2000, and was a collaboration between LBL and SNL-CA. In this campaign, we used resources connected by NTON, a high speed testbed network. For this example, the raw scientific data was located on a DPSS at LBL in Berkeley, while the Visapult back end was located on the CPlant

Linux/Alpha cluster at SNL-CA. The Visapult viewer was running on a desktop machine at SNL-CA. The combustion simulation used for this example was from a 640x256x256 grid, and each grid value was represented with a single IEEE floating point number, for a total of 160 megabytes of data per time step for each of the 265 time steps. The theoretical limit of the network link is 622 Mbps, or the OC-12 connection between LBL and NTON.

**FIGURE 10. NetLogger Instrumentation/Profiling of Visapult**

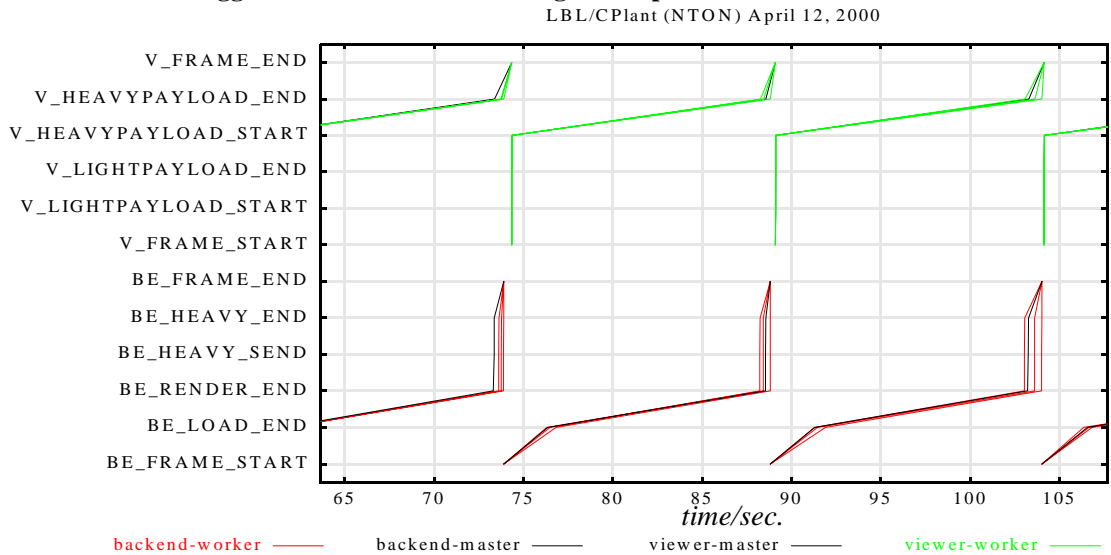

For this image, profile data was collected from both the Visapult back end and viewer. The top row of traces, in green, represent the profile data from the viewer, while the bottom row of traces were obtained from the back end. The horizontal axis represents elapsed time from the start of the application. Each of the entries along the vertical axis of the code are strings associated with specific events, which occurred in order from bottom to top. The viewer events are prefixed with "V_", while the back end events have a "BE_" prefix. Refer to Appendix A for additional details that will aid in interpretation of this data.

In Figure 10, we wish to draw attention to the performance profile of the Visapult back end performance shown by NetLogger instrumentation. The time required to load 160 megabytes of data into the back end from the DPSS over NTON was approximately three seconds[8], for an approximate throughput rate of 433 megabits per second, which is in excess of the network performance realized during the SC99 demonstration over the same network link, reflecting improvements in the underlying Visapult implementation. The improvement in raw network performance was the result of a change to data staging and communications streamlining within Visapult. This amounts to a respectable 70% utilization rate of the theoretical bandwidth limit of the network while data was being transferred. The software rendering then consumed about eight or nine seconds on four processors of the CPlant cluster.

From this campaign, one significant design modification is suggested by the performance data - overlapping network transfers with rendering could have a significant positive impact upon the overall application performance. NetLogger performance profiles, such as that shown in Figure 10, are invaluable for identifying potential performance bottlenecks in distributed applications.

---

8. Displacement along the horizontal axis, time, between the tags BE_FRAME_START and BE_LOAD_END, which bracket the process of moving data from the DPSS into the Visapult back end on CPlant.

## 4.3 Overlapped I/O and Rendering

Each processing element (PE) in the Visapult back end loads a subset of a large scientific dataset, then volume renders it's subset of data. The resulting image is then transmitted to the viewer for use as a two-dimensional texture in a scene graph. Then, the process repeats, looping over time. If loading and rendering were overlapped, so as to occur simultaneously, then we would expect the overall application performance to significantly increase.

**FIGURE 11. Overlapped I/O and Rendering Timing Diagram**

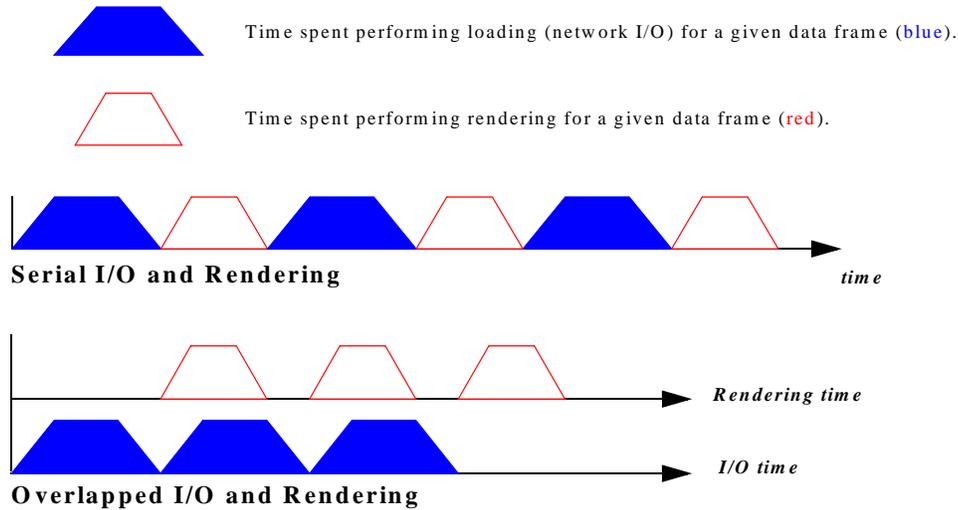

In the discussion that follows, we refer to a *serial* implementation as one in which, in each PE of the parallel Visapult back end, rendering and data loading occur in a serial fashion. Note that even though we use the term *serial*, the back end is in fact a parallel job. Serial refers to how rendering and data loading are executed within each back end process. On the other hand, *overlapped* means that the process of rendering and data loading is implemented in a pipeline-parallel fashion, and occur simultaneously. Also, note that while the data for frame *N* is being rendered, data for frame *N+1* is being loaded.

We can capture the behavior of both serial and overlapped versions, and estimate overall improvement as follows: let *R* be the time spent in each PE performing rendering for each of *N* timesteps of data (the red zones in Figure 11, above), and let *L* be the time spent by each PE loading data for each time step. The amount of time, $T_s$, required for N time steps' worth of data using the serial implementation is: $T_s = N \cdot (L + R)$. In contrast, the time required for *N* time steps using an overlapped implementation is: $T_o = N \cdot max(L, R) + min(L, R)$.

For illustrative purposes, if we assume that *L* and *R* are approximately equal, then the theoretical speedup realized using an overlapped implementation over one that is serial is $T_s/T_o$, or *2N/(N+1)*, which is nearly a 100 percent improvement. As the difference between *L* and *R* increases, the effective speedup resulting from an overlapped implementation will diminish. At one extreme, the overlapped implementation could be as much as nearly twice as fast as a serial implementation. At the other extreme, they will be nearly equal in performance.

The following two figures show the profiling results that compare serial and overlapped implementations of the Visapult back end data loading and rendering tasks. These tests were run using an eight processor Sun Microsystems E4500 server[9] connected to the LBL DPSS via gigabit ethernet (LAN), and were per-formed using ten timesteps from a large scientific data set. The serial implementation required approximately 265 seconds, while the overlapped version required approximately 169 seconds. In each case, *L* was approximately 15 seconds, while *R* was approximately 12 seconds.

**FIGURE 12. Execution Profile of Non-Overlapped I/O and Rendering**

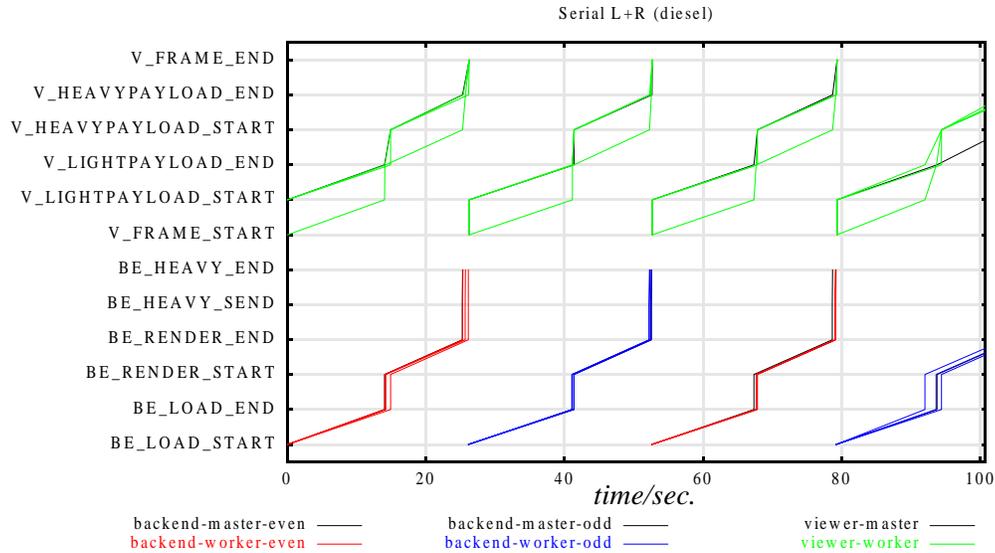

Figures 12 and 13 were created using the NetLogger visualization tool, NLV, and graphically depict and contrast the performance of serial and overlapped implementations of the Visapult back end. In these Figures, the profile traces for the back end are colored according to data frame number; odd frames are blue while even frames are red.

**FIGURE 13. Execution Profile for Overlapped I/O and Rendering**

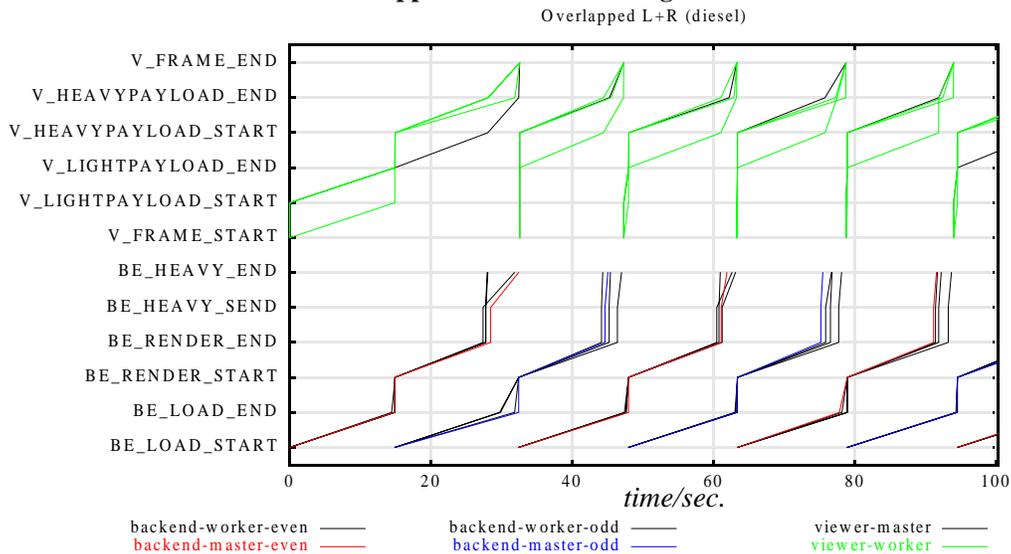

Note that in Figure 13, data loading for frame *N+1* and rendering for frame *N* commence simultaneously. In both the serial and overlapped tests, the time required for each of *L* and *R* are approximately equal. As we shall see in the next section, the time required for each of *L* and *R* in serial and overlapped implementations can vary as a function of the underlying architecture. Details of the overlapped implementation are presented in Appendix B.

---

9. Eight, 336Mhz UltraSparcII processors.

## 4.4 Further Combustion Corridor Testing

In this section, we present performance results obtained while executing Visapult over two different WANs and using two different compute platforms on the back end. One of the WANs, NTON, is a high-speed testbed network that includes an OC12 path from LBL to SNL-CA. The other network, ESnet, is built atop an OC-12 backbone between LBL and ANL, but is a shared resource. The two compute platforms consist of a distributed memory Linux-Alpha cluster, and a large SMP. Each cluster node contains a pair of network interfaces: one for inter-node communication, and the other for external network access. The SMP uses a single gigabit ethernet interface for external network access, which is shared amongst all processors of the SMP. Our goals in the following tests are to obtain an estimate of network bandwidth utilization, and to compare the effect of serial and overlapped implementations of the Visapult back end on two different compute platforms.

### 4.4.1 LBL to CPlant over NTON

In the following two tests, we read data from a DPSS at LBL into CPlant nodes over NTON, performed parallel volume rendering on CPlant, then transmitted the resultant imagery to a viewer at LBL over ESnet. In the earlier campaign that used the LBL DPSS/CPlant/NTON combination (Figure 10), the back end did not yet support overlapped data loading and rendering. The profiles that follow compare and contrast the effect of serial and overlapped data loading and rendering. Figure 14 shows the performance profile of a serial implementation.

**FIGURE 14. Serial L+R on Eight CPlant Nodes**

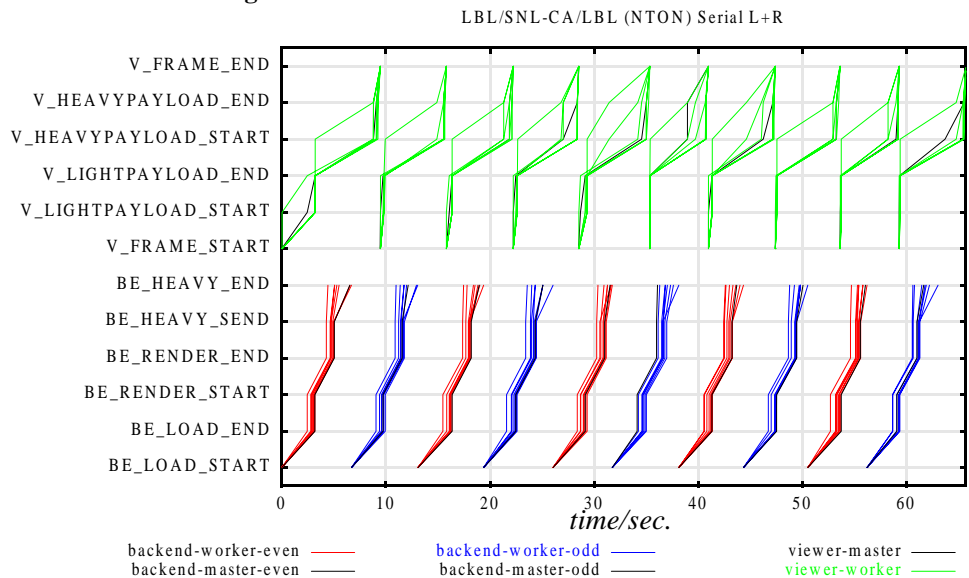

In this example, we used eight nodes of CPlant, a Linux-Alpha cluster. Note that the time required to load 160 MB of data using eight nodes is approximately equal to the time required when using four nodes. From this, we observe that the use of additional nodes will not necessarily improve data throughput, as we have completely consumed all available network bandwidth. On the other hand, rendering time has been reduced to approximately half the time required when using four processors. Given the domain decomposi-tion of the volume data, we expect linear speedup in the rendering process as the number of processors increases.

**FIGURE 15. Overlapped L+R on Eight CPlant Nodes**

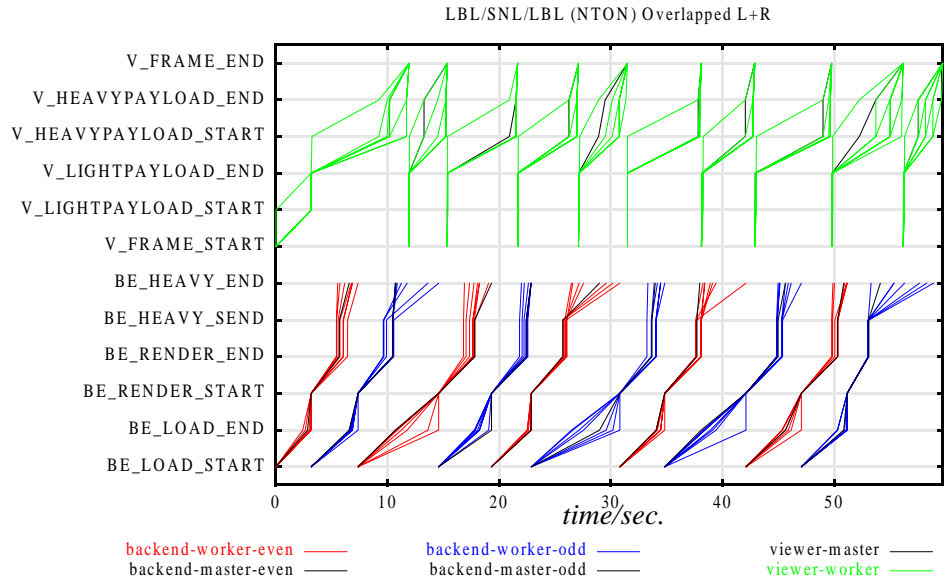

The performance profile in Figure 15 was obtained by running a Visapult back end with overlapped data loading and rendering. One feature in Figure 15 that was expected, but difficult to characterize, is the increased time required for data loading, and the variability in load times from time step to time step. We can presume, based upon the results shown in Figure 15, there may be a relationship between the variability in completion times of transmission of image data from the back end to the viewer and the variability in data loading times[10]. The results indicate that as completion of transmission of outbound images becomes more staggered, inbound data loading is delayed. Another area of interest is CPU contention between the rendering and data loading processes. On CPlant, both rendering and data loading activities share a single CPU. While the render task is CPU intensive and the data loader is an I/O process, there appears to be a significant CPU demand incurred by the data loading process. This may be due in part to implementation details of the underlying network interface card (NIC) driver. It is widely known that some NIC drivers generate more interrupts than others, and these interrupts incur a cost in terms of CPU load. Some gigabit ethernet cards provide the option for using "jumbo frames" (9KB MTUs vs. 1.5KB MTUs), which incur lower interrupt overhead. However, using jumbo frames over a WAN is problematic.

### 4.4.2 LBL to ANL over ESnet

The following two tests contrast serial and overlapped load and render operations on a large symmetric multiprocessing platform with shared memory (SMP)[11] located at ANL. The Visapult back end, running on the SMP at ANL read data from the DPSS at LBL over ESnet, then transmitted partial volume rendering results to a viewer located at LBL, also connected via ESnet. The ESnet link in these tests has a higher latency than the NTON link between LBL and SNL, and delivers an average bandwidth of approximately 100Mbps as measured with commonly available network tools, such as *iperf*[12].

---

10. BE_LOAD_START and BE_LOAD_END bracket movement of data from the DPSS into each back end PE, while BE_HEAVY_SEND and BE_HEAVY_END bracket image transmission from the back end to the viewer.

11. A sixteen processor SGI Onyx2.

Figure 16 shows the performance profile of a serial Visapult back end running on eight processors of the SMP. We observe that approximately ten seconds is required to move 160 megabytes of data per data frame from the DPSS at LBL to ANL over ESnet, yielding a bandwidth consumption of about 128Mbps. Note that data loading time dominates in this case, owing to the significantly lower network capacity. We are able to achieve slightly better bandwidth utilization than a tool like *iperf* owing to the highly parallelized nature of our data loading.

**FIGURE 16. Serial L+R on an SMP**

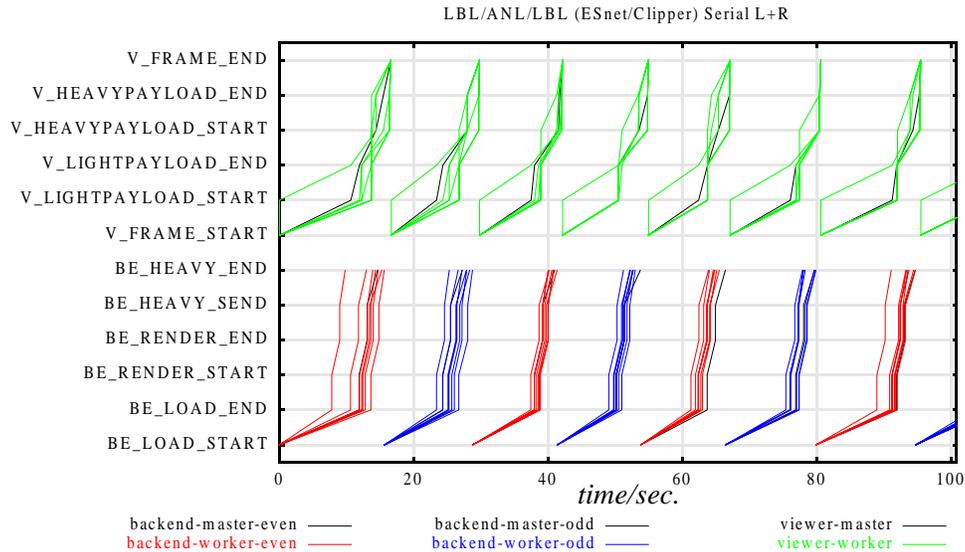

Performance profile for the overlapped Visapult back end is shown in Figure 17. Similar to the NTON/CPlant tests, average elapsed time for overlapped data loading is slightly higher than the serial implementation. After the first time step's worth of data was loaded and the TCP window fully opened, we were able to steadily consume in excess of 100Mbps between the LBL DPSS and ANL over ESnet.

**FIGURE 17. Overlapped L+R on an SMP**

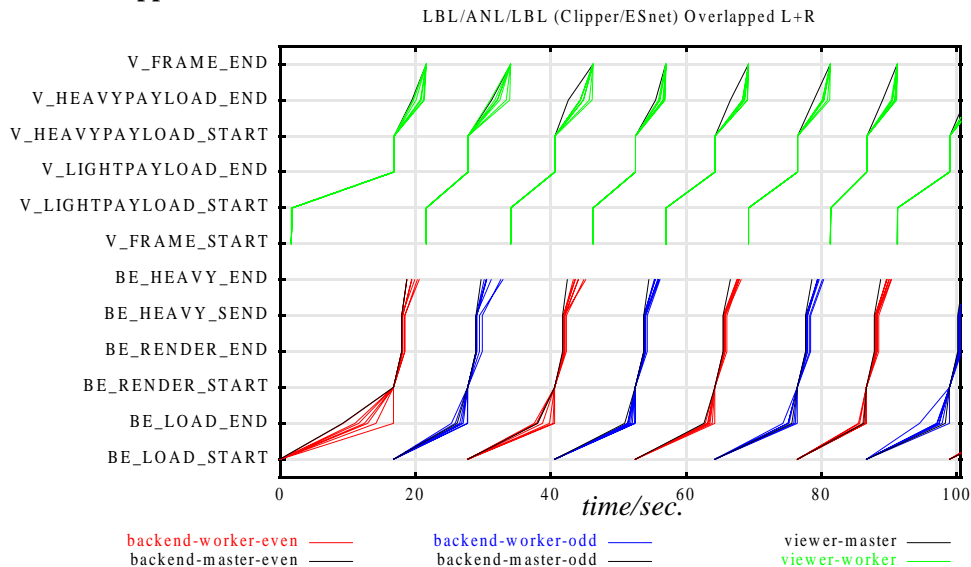

---

12.http://dast.nlanr.net/Projects/Iperf/

It appears that the SMP platform is better suited for the Visapult back end than a distributed memory platform: when each back end process, consisting of a rendering and a data loading thread, map directly onto a CPU, there appears to be less contention and context switching. In contrast, on the cluster, each of the two components of a single back end process must share a single CPU. In addition, the multiple NIC interfaces present on clusters present the possibility of achieving higher aggregate bandwidth utilization than the common SMP configuration of a single NIC shared by all nodes.

## 5.0  Future Work

We have obtained performance numbers for only a subset of contemporary architectures and available high speed testbed networks: large SMPs over a relatively slow and high latency network, and distributed memory systems with single CPU nodes as the compute platform over NTON, a high speed, low latency testbed network. Testing on additional compute platforms, particularly distributed memory architectures with multiple CPUs and shared memory on a single node, as well as an MPI-only implementation of the back end would serve to explore a significant portion of the platform-specific parameter space, and would serve to reveal additional strengths and weaknesses in the overall Visapult architecture.

Access to additional testbed networks is not a trivial task, and often requires the coordination of divergent research and operational groups. From the performance numbers shown in the previous section, it is clear that Visapult completely saturated all of the networks tested, and application throughput will be a function of the capacity of the underlying network. Despite completely using all available network bandwidth, the networks we tested do not have sufficient capacity to meet the challenges of terascale visualization. To put the problem into perspective, the time required to move our 265-timestep dataset (a total of 41.4 gigabytes) over NTON is on the order of eight minutes (a new timestep every 3 seconds), while over ESnet, the time required is on the order of 44 minutes (a new timestep every 10 seconds). A reasonable target rate would be, for this problem, five timesteps per second, requiring effective bandwidth on the order of fifteen times faster than our OC12 connection to NTON; approximately a dedicated OC192 link. This application points out the importance of having Quality of Service (QoS) (including bandwidth reservation) capabilities in future networks. In our testing we were able to completely saturate the WAN link in each network configuration. QoS is needed to insure that this application does not adversely affect other bandwidth-sensitive applications using the link, and to provide some minimum bandwidth guarantees to a Visapult session.

As a parallelized and pipelined implementation of IBRAVR capable of performing interactive volume visualization of large scientific data sets, Visapult's use of IBR-like rendering techniques corroborates the experiences of others who have sought to apply IBR to large model visualization. One such effort used IBR representations of complex geometry as the basis for distance-based model switching as a rendering acceleration aid for navigation through complex CAD models [19]. From a graphics perspective, an architecture built around an embedded scene graph core has proven to be successful in this project. As scene graph technology has been targeted at retained mode rendering of primarily geometric-based data, the question remains as to the applicability of this technology to general IBR techniques. More importantly, the Visapult implementation highlights the relevance of embedded rendering technology within the context of network-based 3D graphics and visualization. Although there are many examples of emerging commercial 3D web-based applications, these tend to use VRML [20] as a medium for data exchange. VRML is a data storage format with an emphasis upon surface and vector geometry. More recently in the VRML97 and Web3D efforts, the VRML base extends geometric modeling to include sound and asynchronous "sensors" that generate events to be consumed and processed by the VRML browser. VRML as a data format doesn't appear to readily lend itself for use by distributed IBR applications: IBR allows for navigation through environments where the source is either precomputed or acquired imagery. We envision interesting future 3D, web-based applications that use the notion of navigating through environments constructed from acquired, rather than computed imagery.

In our experience, remote resource access and management can be a troublesome and tedious endeavor. One of the appealing themes in Corridor projects is the ability of a user to transparently take advantage of remote and distributed resources, such as network storage caches and computational facilities, without specialized knowledge about the distributed resources: access to testbed networks may require modifications to routing tables; the ability to launch a parallel job likely requires shell access to the remote resource; and access to DPSS systems is typically provided on an as-needed basis. In order for research scientists to successfully use a tool like Visapult, they may need detailed technical knowledge of networks, knowledge of the existence of and access to the remote resources, and must be capable of diagnosing the inevitable difficulties that arise when attempting to launch multiple components of a distributed application. Users want tools that are easy to use and help them accomplish their work. A good deal of our future work will be focused upon simplifying the access to and use of the remote and distributed resources upon which Visapult is built.

In this project, the DPSS has proven to be a useful tool. Storage systems of this type present an economical and scalable storage solution that will assume an increasingly important role in a network-centric computing environment. We expect that by augmenting the block data services with additional processing capabilities, the DPSS will become even more useful. For example, "wire level" compression would benefit a wide array of applications. In the case of lossy compression techniques, the degree of lossiness could be a function of network line parameters and under application control. Additional possibilities include off-line visualization services, such as the offline and automatic creation of thumbnail representations of datasets or metadata.

## 6.0  Conclusion

Remote and distributed visualization and rendering algorithms increasingly depend upon a foundation of data management and data movement. As a Corridor project, Visapult has demonstrated the feasibility of using combinations of distributed resources, such as parallel network data caches and computational resources. A unique combination of data staging, parallel rendering and parallel I/O has produced a prototype application and framework that is capable of performing interactive visualization of large scientific data sets. Several instrumented test cases have shown that Visapult is capable of saturating the fastest high speed testbed networks available today. Despite these results, we conclude that these networks are still inadequate for the purposes of tera-scale visualization. Access to the networks can be troublesome, and applications such as Visapult can benefit from related research projects, such as QoS and bandwidth reservation to streamline access to and use of these emerging resources.

## 7.0  Acknowledgement

This work was supported by the Director, Office of Science, Office of Basic Energy Sciences, of the U.S. Department of Energy under Contract No. DE-AC03-76SF00098. Special thanks to Helen Chen, Jim Brandt, Pete Wyckoff, and Mike Hertzer at Sandia National Laboratories for providing access to CPlant and for providing extraordinary support for this project. The scientific data sets used in our experiments were generated by and used with the permission of Julian Borrill, Scientific Computing Group, NERSC and Vince Beckner and John Bell at the Center for Computational Science and Engineering, also at NERSC. Access to computing facilities at Argonne was provided by Rick Stevens and Mike Papka of the Math and Computing Sciences Division at Argonne National Laboratory.

[19] "MMR: An Integrated Massive Model Rendering System Using Geometric and Image-Based Acceleration," D. Aliaga, et. al., in Proceedings of 1999 ACM Symposium on Interactive 3D Graphics.

[20] http://www.vrml.org/

## 9.0 Appendix A - Visapult Internal Architecture

In this Appendix, we provide technical details about the internals of both the Visapult back end and viewer relevant to interpreting the plots of NetLogger profile data.

We begin with a flowchart-like depiction of the Visapult viewer and back end. The flowchart highlights coarse-grained tasks for both the viewer and back end, as well network communication between the cooperative processes.

**FIGURE 18. Visapult Architecture**

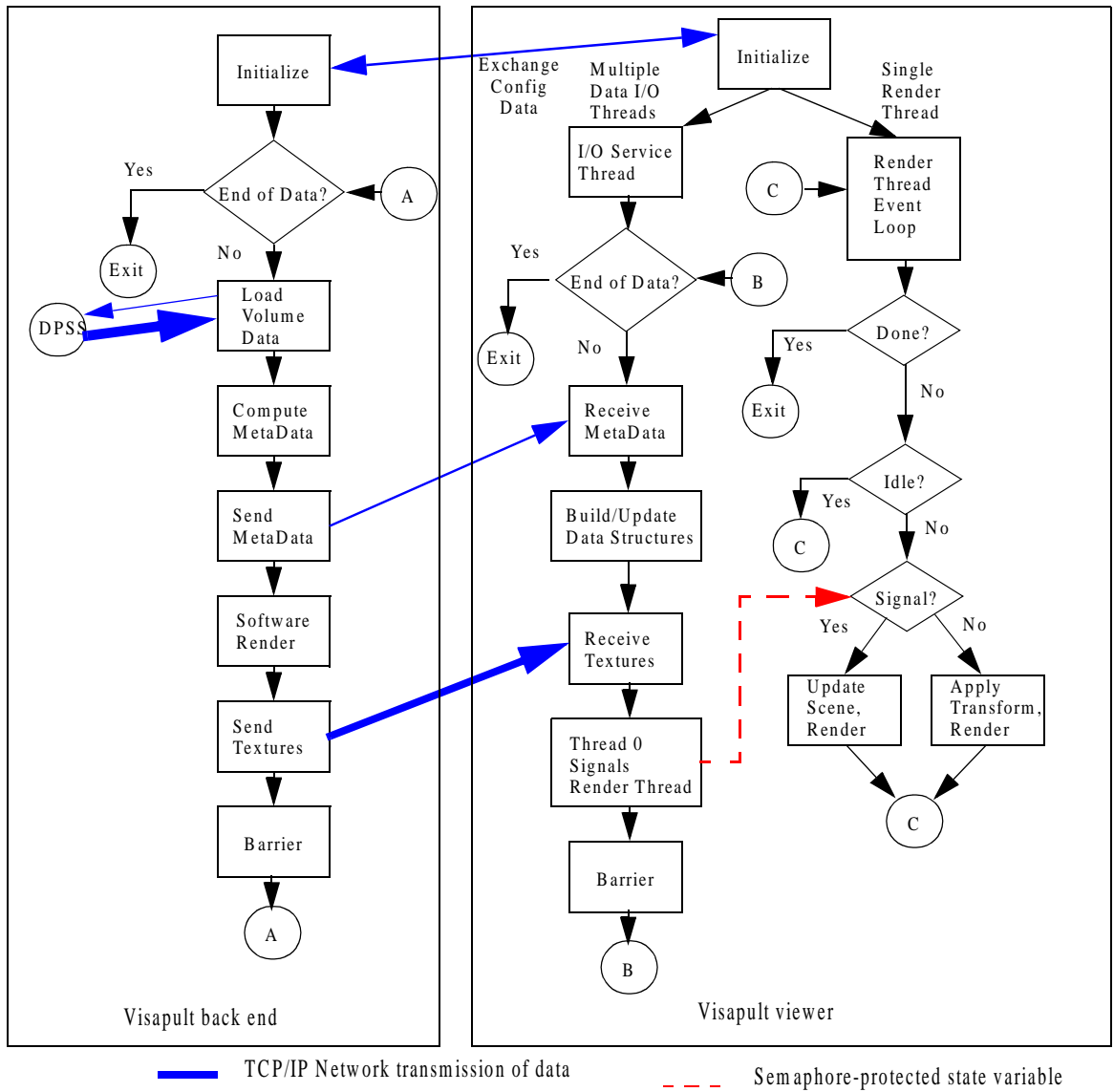

The following two tables provide additional detail about each of the tags present in the profile graphs generated by NetLogger. These tags are used in Figures 10, 12, 13, 14 and 15.

**TABLE 1. Visapult Viewer NetLogger Tags in NLV Figures**

| Tag | Remarks |
| --- | --- |
| V_FRAME_START | Top of loop inside each thread that services an I/O connection with the back end. In the current implementation, the number of time steps, or loops, is set before each of these threads is launched at initialization time. |
| V_LIGHTPAYLOAD_START | Beginning of receipt of visualization metadata. Visualization metadata consists of texture size, bytes per pixel, and geometric information used to place the texture in a 3D scene. Visualization metadata is on the order of 256 bytes, hence the name "light payload." |
| V_LIGHTPAYLOAD_END | Visualization metadata received. |
| V_HEAVYPAYLOAD_START | Beginning of receipt of visualization data. This data consists of raw pixel data, as well as any geometric data, such as triangles, boxes, and so forth. In our tests thus far, the size of this data is also relatively small compared to the size of the source volume. In this implementation, each thread receives a single texture, and while the size of the texture is a function of the resolution of the source volume, a typical size is on the order of 0.25 to 1.0 megabytes per texture. Geometric data is typically tens of kilobytes for the AMR grid data per timestep. |
| V_HEAVYPAYLOAD_END | All visualization data received. |
| V_FRAME_END | End of processing of this time step's worth of data. |

**TABLE 2. Visapult Back End NetLogger Tags in NLV Figures[a]**

| Tag | Remarks |
| --- | --- |
| BE_LOAD_START | Each back end PE is about to load it's subset of volume data. |
| BE_LOAD_END | Volume data load and format conversion completed. In our examples, this step includes loading of AMR species and grid data. |
| BE_LIGHT_SEND | Start transmitting visualization metadata to the viewer. |
| BE_LIGHT_END | Metadata transmission complete. |
| BE_RENDER_START | Start of parallel volume rendering process. |
| BE_RENDER_END | All rendering complete. |
| BE_HEAVY_SEND | Start transmitting visualization data. In this implementation, the visualization data consists of a single texture per back end PE, and optional geometric data representing the grid, and an optional elevation/offset map which the viewer will use to create a quadmesh. |
| BE_HEAVY_END | End of visualization data transmission. |

a. There are many more NetLogger tags present in the Visapult back end. Many were omitted from this table, and from the figures, for brevity. These additional tags are useful for more detailed analysis of execution profiles within each large-grained task (e.g., "load data").

# 10.0  Appendix B - Overlapped Visapult Back End Implementation Details

The Visapult back end is implemented using MPI as the multiprocessing and IPC framework. Each PE of the back end is responsible for reading a subset of the volume data, for rendering its subset of data, and for transmitting the rendering results to the Visapult viewer.

To implement overlapped data loading and rendering in each back end PE, the base MPI code was extended to launch a detached execution thread. We chose to use *pthreads* as the threading API due to its portability and wide availability. In the discussion that follows, we refer to the combination of a single MPI process and its associated detached, reader thread as a *process group* for the sake of clarity. The *reader thread* is the detached, freely-running pthread, and the *render process* is the MPI process. A flowchart of these cooperative processes is shown in Figure 19.

Upon entry, each MPI PE launches a detached, freely-running execution thread (reader thread). This thread logically executes concurrently with the MPI process. *Concurrent logical execution* means that we yield scheduling control to the host system. On distributed memory systems, such as Linux clusters, both reader thread and render process share a single CPU, thereby inviting contention. On SMP systems with a sufficient number of CPUs, in our experience, CPU contention appears to be minimized, if not eliminated.

In our implementation, the reader thread is a worker, and controlled by the render process. Each back end render process creates a pair of SystemV shared memory semaphores prior to launching the reader thread. Each of the semaphore pairs is shared by each render/reader process group, with one such pair for all MPI PEs. One of the semaphores, which we'll call *semaphore A*, is considered as an execution barrier from the perspective of the reader thread, while the other, *semaphore B*, is considered as an execution barrier from the perspective of the render process.

Upon entry to the reader thread, after some internal initialization occurs, the reader thread blocks waiting to gain access to *semaphore A*. The render process will request that either data from a specific time step will be read, or will request reader thread termination due to completion of all time steps. Once the reader thread gains access to *semaphore A*, it will examine the control variable (in shared memory) and take the appropriate course of action, either reading more data or exiting. Upon completion of the requested activity, the reader thread will post to *semaphore B*, then block awaiting access to *semaphore A*.

On the render process side, data from time step zero is first requested from the reader thread. Once that data has been loaded and is available, data from time step one is requested, and the render process begins to render data from time step zero. Once rendering is complete, results are transmitted to the viewer, then the render process will block while attempting to gain access to *semaphore B*. Upon gaining access to *semaphore B*, the render process requests the next time step's worth of data, and posts to *semaphore A*. This process continues until all the time varying data has been read, rendered and results transmitted to the viewer.

In addition to the control semaphores, a large block of memory is shared between reader thread and render process. The reader thread will load the raw scientific data into this large memory block during reading. This memory is considered to be *double-buffered*: its size is twice that of a single time step's worth of data, and the reader thread will use one half of the buffer for writing into, while the render process reads from the other half. Access control is implicit as a function of the time step using an even-odd decomposition. Due to the control architecture of the reader thread and render process, we are guaranteed that reader and render threads will not access the same odd/even data buffer at the same time.

We chose to extend the MPI base using pthreads in order to take advantage of the shared-memory model employed by threaded code. An alternative would be to use MPI-only constructs. For example, even-num-bered processes would render, while odd-numbered processes would read data. The synchronization between the two would be similar, but using MPI constructs rather than SystemV semaphores. Of greater concern would be the need to transmit large amounts of scientific data between reader and render processes. We consciously chose to avoid incurring this additional cost by using a threaded model. In doing so, we may have incurred a penalty in the form of increased contention on distributed memory architectures with single-CPU nodes.

**FIGURE 19. Architecture of Overlapped Visapult Back End**

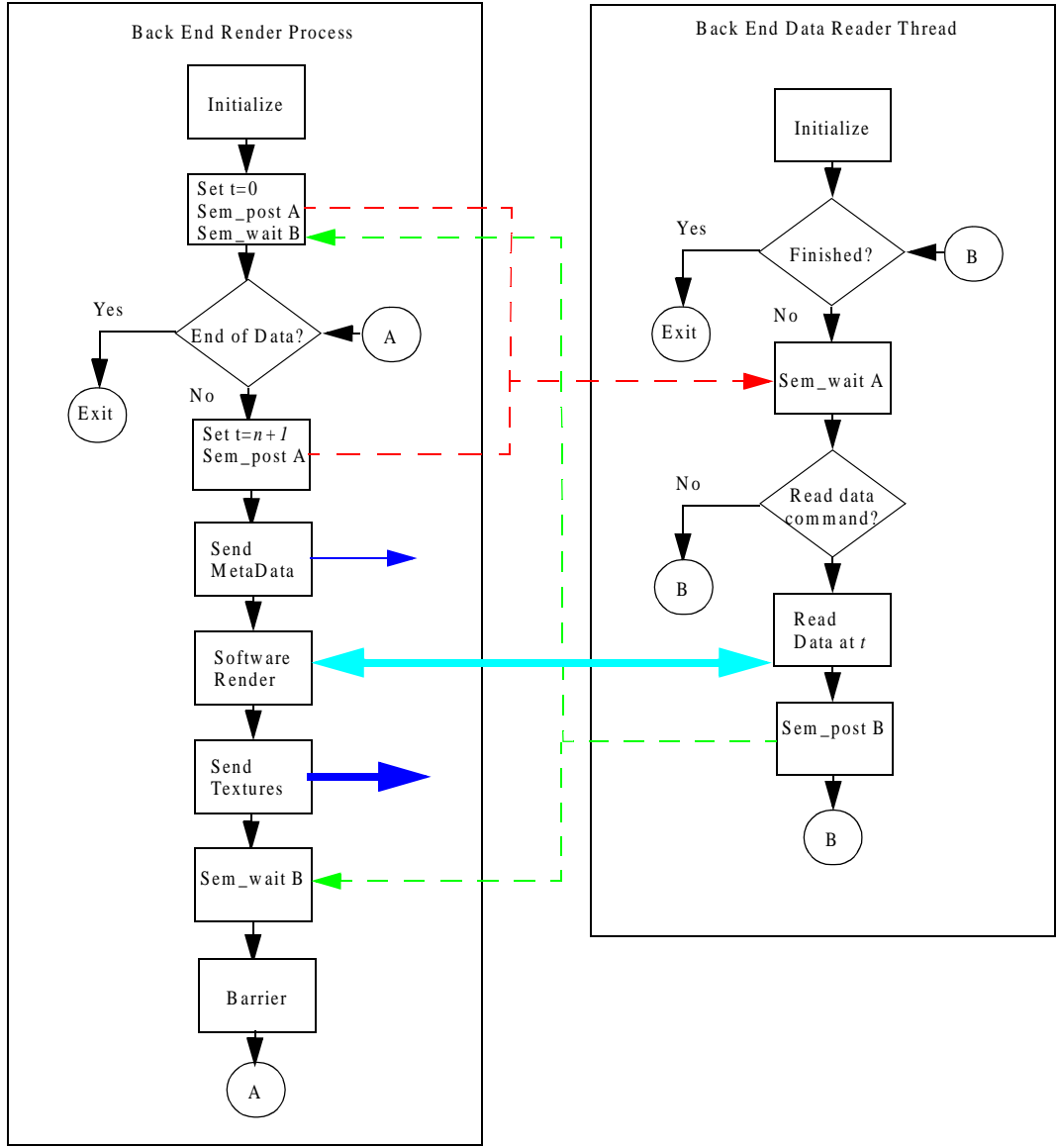